*Mini Review*

# Sensitivity Analysis and Parametric Optimization of Micro-Plasma Actuators: A Mini Review

**Javad Omidi [1*]**

[1] Research Assistant, Department of Chemical Engineering, Columbia University, NYC, USA
\* Correspondence: jo2668@columbia.edu

**Abstract:** The Dielectric Barrier Discharge (DBD) micro-plasma actuator stands out as a highly promising tool for active fluid flow control. Researchers specializing in flow control have taken a keen interest in this actuator due to its economical manufacturing, low energy consumption, compact size, lightweight nature, straightforward implementation, and absence of movable components or pneumatic/hydraulic systems. Given its extensive application, achieving the best design for plasma actuators necessitates a more profound grasp of how diverse physical factors (like electrode thickness, electrode length, dielectric thickness, and dielectric materials) and operational variables (such as applied voltage, frequency, and waveform) impact its performance. Within this article, we delve into a comprehensive assessment of both numerical and experimental investigations focused on optimizing actuator parameters. These studies can be categorized into two main groups. The initial group involves fundamental test cases conducted on flat plates, while the subsequent group pertains to modeling controlled flow in real-world scenarios, including curved surfaces.

## 1. Introduction

To produce plasma in atmospheric conditions, Dielectric-barrier-discharge (DBD) plasma actuators are known to be the least expensive and the most productive generators which can be used in various applications [1 - 8]. The main discharge phenomenon was discovered in 1857, but studies on them as actuators returns back to much later, as 1987 [9], and their first usage as a flow control device is due to Roth [10], as recently as 1998. These actuators have been of great interest for researchers in flow control devices, because of their low production and application expenses, low power usage, small dimensions, low weight, easy implementation, and that they have no moving parts or pneumatic or hydraulic systems. Recently they have been studied by different researchers for different applications [11 - 16]. Electro-hydrodynamic plasma actuators have proven to be effective for flow attachment in internal and external aerodynamics, and for improvement of the lift, drag, and stall angle of airfoils. The performance of plasma actuators has been studied with such classical aerodynamic tools as wind tunnels, drag balances, Pitot tubes, smoke flow visualization, and fluid dynamic modeling programs. However, the physical processes and power flows that occur in plasma actuators, before that the plasma ions transfer their momentum to the neutral background gas are mostly an electrical process. For wind turbine applications, the interaction of this plasma region with the boundary layer is very important. Design and optimization of such actuators needs to combine the methods of electrical engineering and plasma physics, including classical electrical discharge physics, with flow hydrodynamics.

Due to their extensive usage in different applications, we need to a better understanding of the effect of different geometrical or performance parameters (e.g. the applied voltage, frequency, and the waveform). Ultimately, we would like to optimize design of these actuators for different applications. The numerical and experimental studies on optimization of different parameters of these actuators may be categorized to two groups: 1) application on a flat plate, 2) usage on a curved surface for a real application.

Most of these studies are experimental, and to have an accurate real experimental model, usually requires extensive time and resources, and limitations in measurement devices usually restrict experimentations in full scales. On the other hand, new achievements in numerical analysis and exponential increase in the computational power, has made it feasible to have accurate enough simulations of the flow field and flow control devices. In the past, due to deficiencies of plasma simulations models, and their weakness in simulation of interactions of the plasma with the flow field, most parameter or optimization studies are performed experimentally. Here we review first a few of these studies for parametric or optimization studies of DBD actuators for generation of ionized wind over a flat plate. Afterwards, we will review some recent works for application f DBD actuators on curved surfaces.

**2. Flat Plate and Basic Test Cases Investigations**

*2.1. Experimental Studies*

Forte et al. [17] performed an experimental parametric study in order to increase the velocity of the ionic wind induced by such actuators. The results show that an optimization of geometrical and electrical parameters allows us to obtain a time-averaged ionic wind velocity up to 8 m/s at 0.5 mm from the wall.

In another experimental optimization activity, Abe et al. [18] investigated optimization of the momentum transfer performance of the plasma actuator for a variety of parameters, including the ambient gas pressure, ambient gas species, alternating high voltage wave form, electrode configuration and dielectric plate material. It was found that the ambient gas pressure, under which a plasma actuator operates, has a significant effect on the momentum transfer performance.

In an experimental work, Robert Van Dyken et al. [19] studied several parameters on a flat plate substrate in still air and validated the results on a thin airfoil at low speeds. These parameters included the signal waveform, signal frequency, electrode geometry, as well as power dissipated by the plasma generating system, and showed that an optimized plasma actuator gave substantial increases in the lift and reductions in the drag beyond the baseline stall angle of attack.

To optimize the body force produced by a single DBD actuator for aerodynamic flow control, Thomas et al. [20] performed an experimental investigation. A primary goal of the study was to improve the actuator authority for flow control applications at higher Reynolds number than previously possible. The study examined the effects of the dielectric material and thickness, applied voltage amplitude and frequency, voltage waveform, exposed electrode geometry, covered electrode width, and multiple actuator arrays. It was demonstrated that actuators constructed with thick dielectric material of low dielectric constant produce a body force that is an order of magnitude larger than that obtained by the Kapton-based actuators used in many previous plasma flow control studies. These actuators allow operation at much higher applied voltages without the formation of discrete streamers which lead to body force saturation.

The performance of arrays of plasma stream-wise vortex generators (PSVG) was experimentally compared with passive VGs by Wicks et al. [21]. The arrays were flush mounted on a flat plate with a turbulent boundary layer. They considered the influence of freestream velocity, applied peak-to-peak voltage, length of the covered electrode and span-wise inter-electrode spacing, on actuator performance. Results demonstrated that the performance of PSVG is similar to a passive VG. Similarly, Jukes et al. [22] investigated the use of DBD plasma actuators as vortex generators for flow separation control applications. Plasma actuators were placed at a yaw angle to the oncoming flow, in two co-rotating and counter-rotating vortex arrays so that they produced a span-wise wall jet. Through interaction with the oncoming boundary layer, this created a stream-wise longitudinal vortex. In this experimental investigation, the effect of yaw angle, actuator length and plasma-induced velocity ratio was studied. The vortex generators were successful in reducing the separation region, even at plasma-to-freestream velocity ratios of less than 10%.

Erfani et al. [23] exploited multi-encapsulated electrodes to produce higher velocities providing more momentum into the background air. As the number of encapsulated electrodes increased and variation of other variables such as the driving frequency and voltage amplitude were considered, they found the optimum actuator configuration. They designed a handful of experiments, for which the velocity is obtained by Particle Imaging Velocimetry measurement. They also used surrogate modeling to find the velocity, and the model was validated both experimentally and statistically. Then, numerical optimization was conducted and results were investigated through experiment. The results showed that the surrogate modelling approach provides a cheap and yet efficient method for systematically investigating the effect of different parameters on the performance of the plasma actuator.

In an experimental work, Neretti et al. [24] reported on the importance of different electrode geometries on the performance of different Plasma Synthetic Jets Actuators. A series of DBD aerodynamic actuators designed to produce perpendicular jets. Linear and annular geometries were considered, and optimal values of upper electrode distances in the linear case and different diameters in the annular one were found, to maximize jet velocity, mechanical power or efficiency. Annular geometries were found to achieve the best performances.

Taleghani et al. [25] performed a parametric experimental study to optimize the geometrical and electrical parameters to increase the velocity of the ionic wind. Results showed that increments in the excitation frequency, leads to a higher velocity at a place close to the surface. The frequency of the vortices produced by the actuators is the same as the excitation frequency and an increase in the duty cycle increases the power of the produced vortex shedding.

*2.2. Numerical Studies*

Yoon and Han [26] presented an improved model by replacing the empirical formulae in their previous model with physical equations that take into account physical phenomena and environmental variables. Additional operation parameters, such as pressure, temperature and ac waveforms, were taken to predict the thrust performance of the actuators with a wider range of existing parameters, the thickness of the dielectric barrier, the exposed electrode, the dielectric constant, the ac frequency and the voltage amplitude.

Experimental studies have reported that a thick dielectric (a few mm) enhances the plasma extension, resulting in an improvement of the electro-hydrodynamic (EHD) body force and the resulting ionic wind. Various parameters influence the electro-mechanical characteristics of DBD actuators such as the dielectric thickness, the distance between electrodes, the amplitude, the frequency and the waveform of the applied voltage. Seth et al. [27] extended the Suzen and Huang [28 and 29] model to analyze the effects of some of these geometrical and electrical parameters of DBD plasma actuators in flow control applications.

In a numerical optimization, Regis de Quadros et al. [30] studied active wave cancellation for a flat plate boundary layer, with an adverse pressure gradient at low Reynolds number, using a pulsed plasma actuator. They focused on the linear stage of the growth of disturbances, where delaying the laminar-turbulent transition by an active control is more easily achievable. They also considered the algorithm of Nelder and Mead as the most popular simplex method in practice for unconstrained optimization. The impact of fluctuating and transient kinematic and thermodynamic airflow conditions on the performance of dielectric barrier discharge (DBD) plasma actuators were demonstrated by Kriegseis et al. [31]. A novel online-characterization and control approach was introduced, revealing the possibility of compensating for impaired discharge performance due to changing airflow scenarios during actuator operation. The goal of online and in situ controlling of the plasma actuator performance was achieved and successfully demonstrated.

Benard et al. [32] coupled a numerical genetic algorithm optimizer with an experiment to minimize the reattachment point downstream of the backward-facing step model. Coupled with the DBD plasma actuator and the wall pressure sensors, the genetic algorithm finds the optimum forcing conditions to reduce the mean flow reattachment by 20%. This is achieved by forcing the flow at the shear layer mode where a large spreading rate is obtained by increasing the periodicity of the vortex street and by enhancing the vortex pairing phenomena. In another work, Benard et al. [33] used an autonomous multi-variable genetic single-objective optimization in an experiment for simultaneous optimization of the voltage amplitude, the burst frequency and the duty-cycle of the high voltage signal. The genetic algorithm can find the optimum forcing conditions in only a few generations. In another numerical/experimental work, Benard et al. [34] proposed a numerical method to estimate the EHD force by using experimental velocity information. They used a parametric study and the obtained results were compared with force balance measurements. It was shown that the EHD volume force increased in space and in amplitude by increasing the voltage and the AC frequency.

## 3. Curved Surfaces and Real Applications

*3.1. Experimental Studies*

Now we review a few of recent numerical or experimental investigations for parametric or optimization studies, implemented on curved surfaces. Greenblatt and Wygnanski [35] performed a parametric study to investigate the effect of periodic excitation (with zero net mass flux) on a NACA0015 airfoil under incompressible flow conditions, undergoing pitch oscillations at rotorcraft reduced frequencies, to maximize the airfoil performance while limiting moment excursions to typical pre-stalled conditions. Significant increases in maximum lift and reductions in drag were attained. Oscillatory excitation was found to be far superior to steady blowing, which was even detrimental under certain conditions, and flap-shoulder excitations were found to be superior to leading-edge excitations.

Roth and Dai [36] implemented a program of optimization by decomposing the power flow through a plasma actuator into four sinks: 1) Reactive power losses due to inadequate impedance matching of the power supply to the actuator; 2) Dielectric heating of the actuator insulating materials; 3) Power required to maintain the atmospheric pressure plasma; and 4) Power coupled to the neutral gas flow by ion-neutral collisions. These four power flows can be, and usually are, of comparable magnitude. They reviewed the progress in understanding and minimizing the first three power flows, and maximizing the fourth by adjustment of the actuator geometry and materials, as well as such plasma parameters as the RF frequency and RMS voltage.

In an experimental work on a NACA 0015 airfoil, Jolibois et al. [37] used a DBD to modify velocity in the boundary layer. The goal of the actuation is to displace (upstream or downstream) the separation location, in either reattaching a naturally detached airflow or in detaching a naturally attached airflow. They showed that the plasma actuator is more effective when it acts close to the natural separation location, and that the power consumption can be highly reduced in using a non-stationary actuation.

In another experimental work on micro thrusters and internal duct aerodynamics, Ozturk et al. [38] used a closed circumferential arrangement to yield a body force. The primary flow is driven by this zero-net mass flux jet at the wall that then entrains fluid in the core of the duct. This resulted in a unique configuration for studying impulsively started jet phenomena. Several experiments were conducted on tubes of different diameters under varying parameters such as the modulation frequency and the duty cycle. They found higher induced velocities increase with higher forcing frequencies and duty cycles, although there is a peak value for the forcing frequency, after which the velocity and thrust decrease. The influence of the length-to-diameter ratio was also significant; the velocities and thrust increase as the inner diameter of the tubes are increased. Velocity profiles show a great difference with this ratio; the smallest diameters yield a peakier profile whereas larger diameters produce a broader profile with low centerline velocities.

Models for the time and space dependence of the body force on the input voltage amplitude, frequency, electrode geometry, and dielectric properties have been developed and used, along with experiments, to optimize the actuator performance. Corke et al. [39] have overviewed some of applications that include leading-edge separation control on airfoils, dynamic-stall vortex control on oscillating airfoils, and trailing-edge separation control on simulated turbine blades to highlight the plasma actuator characteristics and modeling approach.

Using an optimization algorithm using multiple input parameters, Matsuno et al. [40] carried out the optimization of the driving conditions of the plasma actuators by using a robust design method for wake control at high dynamic pressure conditions. This resulted in optimal condition for reduced control parameters. In another work, Matsuno et al. [41] applied a Kriging–based genetic algorithm called efficient global optimization (EGO). The aerodynamic performance was evaluated by wind tunnel testing for a drag minimization problem around a semicircular cylinder, to find the optimal pulse drive conditions for plasma actuators.

Sulaiman et al. [42] experimentally used a Multi-Objective Design Optimization method to optimize a DBD actuator over an airfoil. Experiments were conducted at a Reynolds number of 63,000 using a NACA0015 airfoil fixed to the stall angle of 12 degrees. Minimization of the power consumption and maximization of the lift coefficient were used as objective functions. The algorithm was converged with only 10 generations with a total population of 260 solutions.

Batlle et al. [43] described and experimentally evaluated an airfoil design methodology, in which airfoils are designed to be simultaneously suited for wind energy applications and to employ DBD plasma actuators as AFC (active flow control) devices. A multi-objective optimizer is used with a genetic algorithm, which evaluates 150 airfoil candidates along 30 generations. Two cost functions are introduced. The wind energy suitability cost function compares the non-actuated airfoil performance with a reference group of airfoils. The AFC suitability cost function assesses the changes in the CL (named CL authority) caused by the DBD actuation. The parametric study includes different wind speeds, DBD voltage amplitude and actuation direction. Higher voltages applied do not lead to a significant increase of the CL authority and different wind speeds and actuation direction reduce the actuation efficiency.

*3.2. Numerical Studies*

In an extensive numerical simulation, Sato et al. [44] used a massive number of large-eddy simulations of the separated flow over NACA0015 airfoil, controlled by a DBD plasma actuator to reduce power consumption. Study parameters included the position and operation conditions of the DBD plasma actuator, (e.g. the burst frequency, the degree of induced flow and the burst ratio of actuation). They verified that the most effective position of the actuator to suppress the separation is the vicinity of the separation point. The most effective burst frequency of burst wave to improve the lift-drag ratio was F =5.

Watanabe et al. [45] presented an algorithm for multi-objective evolutionary optimization, which was based on the NSGA-II with the Chebyshev preference relation. The scheme was applied to a multi-objective design optimization problem of dielectric barrier discharge plasma actuator (DBDPA). This optimization problem had four design parameters and six objective functions. The main goal of the paper was to extract useful design guidelines to predict control flow behavior based on the DBDPA parameter values using the resulting approximate Pareto set obtained by the optimization algorithm.

In a numerical simulation conducted by Williams et al. [46], a Quantitative Design Optimization approach was presented for active flow control using DBD plasma actuators. The approach couples geometric changes (including aerodynamic shape optimization make using the adjoint formulation of the Navier-Stokes equations for incompressible flows) and plasma actuator design to find optimum flow control conditions. Ultimately, the optimization seeks a given set of merit functions including a required lift control with minimum actuator power.

The location of the plasma actuator in a chord-wise direction on a wind turbine airfoil is numerically investigated and studied by the authors [2, 5, and 11]. The results show that the actuators which are close to the separation point have a more efficient effect on flow separation control. In these works also the single and tandem actuation is analyzed which showed that the tandem actuation would work better. The authors [2] studied two operational parameters of applied voltage and frequency. In this regard one of the most phenomena was that after a certain amount of frequency, the effect of increasing decrease to certain point.

The authors, in another work [47], investigated a parametric study of the micro-plasma actuator installed on a wind turbine airfoil in a stall condition to reduce and control the size of the fluid flow separation. In this regard the geometrical parameters and dielectric materials are studied. Initial increase of the length of the embedded electrode improves the aerodynamic performance of the airfoil, and after reaching a certain amount, the lift coefficient slightly decreases, while the drag coefficient continues to decrease. Increasing the dielectric permittivity is also very effective in improving the aerodynamic performance up to a certain value, and after that more increase in the permittivity coefficient has little effect. The effect of the dielectric thickness on the aerodynamic improvement was not favorable, and a fairly fast reduction in the performance was observed, recommending the least possible dielectric thickness. Also, increasing the thickness of the electrodes reduces the aerodynamic performance, and this is also more pronounced up to a thickness of a certain amount, after which the plasma actuator has little effect on the aerodynamic performance of the blade. The influence of operational factors such as applied voltage, frequency, and wave form is also thoroughly investigated in another study [48]. The author [49] conducts an optimization research in order to improve the implementation and optimal configuration of plasma actuator in wind turbine airfoils. Its findings yielded a formula for each of the best geometrical and material characteristics for each voltage and frequency combination.

## 4. Conclusions

One of the most challenging aspects of this type of actuator which could be shown in this review is that the computational works are limited in comparison to the experimental ones. Actually, computational analysis of the real physics of DBD actuators is very complicated, involving interactions of ionization, fluid flow and the electrical field. Accurate solution requires simultaneous solution of the Maxwell and Navier-Stokes equations. There are few attempts to solve this complicated non-linear combination [1 - 3, 6, 23, 24, and 50 - 52], but it is practically assumed to be too expensive to be used. The lack of an effective numerical model in this field is evident. This is better observed in the design and optimization phases of plasma actuator studies. To have a comprehensive parametric (both ge-ometrical and electrical) study on application of a DBD plasma actuator for different applications requires availability of an appropriate model with little cost and suitable accuracy.